\documentstyle[amssymb,preprint,prd,aps]{revtex}

\global\arraycolsep=2pt
\makeatletter
\def\fmslash{\@ifnextchar[{\fmsl@sh}{\fmsl@sh[0mu]}}
\def\fmsl@sh[#1]#2{  \mathchoice
    {\@fmsl@sh\displaystyle{#1}{#2}}    {\@fmsl@sh\textstyle{#1}{#2}}    
{\@fmsl@sh\scriptstyle{#1}{#2}}    {\@fmsl@sh\scriptscriptstyle{#1}{#2}}}
\def\@fmsl@sh#1#2#3{\m@th\ooalign{$\hfil#1\mkern#2/\hfil$\crcr$#1#3$}}
\makeatother

\begin{document}
\draft
\title{Distinguishing locally of quantum states and the distillation of entanglement}
\author{Ping-Xing Chen$^{1,2}${\footnotesize \thanks{%
E-mail: pxchen@nudt.edu.cn}\thanks{%
present address: address 2}} and Cheng-Zu Li$^2$}
\address{1. Laboratory of Quantum Communication and Quantum Computation, \\
University of Science and Technology of\\
China, Hefei, 230026, P. R. China \\
2. Department of Applied Physics, National University of\\
Defense Technology,\\
Changsha, 410073, \\
P. R. China. }
\date{\today}
\maketitle

\begin{abstract}
This paper try to probe the relation of distinguishing locally and
distillation of entanglement. The distinguishing information (DI) and the
maximal distinguishing information (MDI) of a set of pure states are
defined. The interpretation of distillation of entanglement in term of
information is given. The relation between the maximal distinguishing
information and distillable entanglement is gained. As a application of this
relation the distillable entanglement of Bell-diagonal states is present.
\end{abstract}


\thispagestyle{empty}

\newpage \pagenumbering{arabic} 

A basic features of quantum mechanics is that one cannot distinguish a set
of non-orthogonal states. If the states are shared by bipartite (Alice and
Bob) and only local operation and classical communication (LOCC) are
allowed, in general, one even cannot distinguish a set of orthogonal states 
\cite{1,2}. Some interesting works on locally distinguishability of quantum
states have been present\cite{1,2,3,4,5}. For example, any three of the four
Bell states

\begin{eqnarray}
\left| \Phi ^{\pm }\right\rangle &=&\frac 1{\sqrt{2}}(\left| 00\right\rangle
\pm \left| 11\right\rangle )  \label{1} \\
\left| \Psi ^{\pm }\right\rangle &=&\frac 1{\sqrt{2}}(\left| 01\right\rangle
\pm \left| 10\right\rangle )  \nonumber
\end{eqnarray}
cannot be distinguished by LOCC operation if only a single copy is provided 
\cite{2}. Another basic features of quantum mechanics is entanglement. On
the hand, maximally entangled states may have many applications in quantum
information, such as error correcting code\cite{6}, dense coding\cite{7} and
teleportation\cite{8}, etc. On the other hand, in the laboratory, however,
maximally entangled state became a mixed state easily due to the interaction
with environment. This results in poor application. The idea of distillation
protocols is to get some maximally entangled states from many or infinite
copies of a mixed states. A few of distillation protocols was given\cite{9},
but finding the most efficient distillation protocol and calculation of
distillable entanglement (the maximal value of entanglement gained from per
mixed state), $E_D,$ are still open questions. All distillation protocols
have a common feature: the distillable entanglement from a mixed state is
not more than the entanglement of formation of the mixed state owing to the
loss of information\cite{10,11}. In essence, indistinguishability of a set
of orthogonal entangled states is also owing to the loss of information. The
transformation of information acts as a important role in both distillation
of entanglement and distinguishing of states. In this sense, the
distinguishing locally and the distillation of entanglement should have some
connection. In this paper, we try to probe this question. First we define
the distinguishing information (DI) and the maximal distinguishing
information (MDI) of a set of pure states (which are, in general, entangled
states). Then we give an interpretation of distillation of entanglement in
term of information and get the relation between MDI and distillable
entanglement $E_D$. Finally as a application of this interpretation and this
relation we discuss the MDI and get the $E_D$ of Bell-diagonal states

\begin{equation}
\rho =p_1\left| \Phi ^{+}\right\rangle \left\langle \Phi ^{+}\right|
+p_2\left| \Phi ^{-}\right\rangle \left\langle \Phi ^{-}\right| +p_3\left|
\Psi ^{+}\right\rangle \left\langle \Psi ^{+}\right| +p_4\left| \Psi
^{-}\right\rangle \left\langle \Psi ^{-}\right|  \label{2}
\end{equation}

Locally distinguishing, in essence, consists of local unitary transformation
and local measurement with the help of classical communication. Suppose
there are a set of pure bipartite states:

\begin{equation}
\left| \Psi _1\right\rangle ,\left| \Psi _2\right\rangle ,...,\left| \Psi
_i\right\rangle ,...,\left| \Psi _n\right\rangle \quad \quad \quad \quad
i=1,...,n
\end{equation}
which are shared by Alice and Bob. In general, one cannot distinguish
deterministically each state, but it is possible that after locally
distinguishing operation one can determine that a state is one of a few of
possible states and not the other states. It is to say one can divide n
states into a few of groups states by distinguishing locally. For example,
by the use of a complete set of projecting operation:

\begin{equation}
\left| 00\right\rangle \left\langle 00\right| ,\left| 01\right\rangle
\left\langle 01\right| ,\left| 10\right\rangle \left\langle 10\right|
,\left| 11\right\rangle \left\langle 11\right|   \label{44}
\end{equation}
one can distinguish $\left| \Phi ^{\pm }\right\rangle $ from $\left| \Psi
^{\pm }\right\rangle $ in four Bell-diagonal states (but cannot distinguish $%
\left| \Phi ^{+}\right\rangle $ from $\left| \Phi ^{-}\right\rangle $ or $%
\left| \Psi ^{+}\right\rangle $ from $\left| \Psi ^{-}\right\rangle )$,
i.e., divide four Bell-diagonal states into two group states $\left| \Phi
^{\pm }\right\rangle $ and $\left| \Psi ^{\pm }\right\rangle .$ If one can
divide locally n states into k groups states with corresponding probability $%
p_k$, one can get the information of these pure states, $I_d$

\begin{equation}
I_d=-\sum_{i=1}^kp_k\ln p_k  \label{4}
\end{equation}
We define $I_d$ as distinguishing information (DI). For Bell-diagonal states
and the projecting operation above$,$ one can get $\left| \Phi \right\rangle 
$ with probability $p_1+p_2$ and $\left| \Psi \right\rangle $with
probability $p_3+p_4.$ So the distinguishing information is

\begin{equation}
I_d=-(p_1+p_2)\ln (p_1+p_2)-(p_3+p_4)\ln (p_3+p_4)  \label{5}
\end{equation}
Different local operation may result in different DI, but for a definite set
of states there is the maximal distinguishing information (MDI), $I_{d\max
}. $ In the latter of this paper we will see that the MDI and the
distillable entanglement have close relation.

Now we discuss the relation between distinguishing locally and the
distillation. A mixed state $\sigma $

\begin{equation}
\sigma =\sum_i\lambda _i\left| \Phi _i\right\rangle \left\langle \Phi
_i\right| ,\quad \sum_i\lambda _i=1  \label{7}
\end{equation}
is shared by Alice and Bob. The distillation protocol is to get some
maximally entangled states (or some pure entangled states because pure
entangled states can be transferred into maximally entangled states
reversibly) from many or infinite copies of a mixed entangled states $\sigma
,\sigma ^{\otimes n}$

In term of information, $\sigma ^{\otimes n},$ which have $2^{nS(\sigma )}$
''likely'' strings of orthogonal pure states, such as, one of strings

\[
\stackrel{n-3}{\overbrace{\left| \Phi _1\right\rangle \cdots \left| \Phi
_1\right\rangle }}\left| \Phi _2\right\rangle \left| \Phi _3\right\rangle
\left| \Phi _4\right\rangle , 
\]
include $nS(\sigma )$ bits information. Where $S(\sigma )$ is the
information entropy of $\sigma $

\begin{equation}
S(\sigma )=-\sum_{i=1}\lambda _i\ln \lambda _i  \label{8}
\end{equation}
One cannot distinguish the $2^{nS(\sigma )}$ strings without destroying the
entanglement of each string owing to the irreversibility of mixing process,
but may distinguish the $2^{nS(\sigma )}$ strings, i.e., get $nS(\sigma )$
bits information by the measurement on some pairs, the entanglement of which
would be lost unavoidably. The essence of distillation is to distinguish the 
$2^{nS(\sigma )}$ strings by measurement on some pairs. The more the
measured pairs to distinguish the $2^{nS(\sigma )}$ strings are, the lesser
the distillable entanglement of per pair is. The maximal distillable
entanglement of per pair is defined as $E_D$ of the mixed state $\sigma $ 
\cite{6,12}. Suppose one need at least measure ($n-m)$ pairs particles to
get $nS(\sigma )$ bits information. To get $nS(\sigma )$ bits information by
the measurement on the ($n-m)$ pairs particles, one may apply the local
unitary transformation to the all copies $\sigma ^{\otimes n}$ so that when
one measure the left (or right) ($n-m)$ pairs particles, one can get the MDI
of each pair particles, $I_{d\max }(\sigma )$. From the following of this
paper one can find easily that if n is infinite, one can get same MDI from
any one of ($n-m)$ pairs particles, so the MDI from ($n-m)$ pairs particles
is $(n-m)I_{d\max }(\sigma )$. When equality 
\begin{equation}
nS(\sigma )=(n-m)I_{d\max }(\sigma )  \label{9}
\end{equation}
holds one can get the information from $(n-m)$ pair particles as much as
that $\sigma ^{\otimes n}$ includes, i.e., one can distinguish the $%
2^{nS(\sigma )}$ strings in $\sigma ^{\otimes n}$ at least expense of
measured pairs. Because each of the $2^{nS(\sigma )}$ ''likely'' strings has
same entanglement, after measurement on the left ($n-m)$ pairs particles in
each string the right unmeasured $m$ pairs have also same entanglement if $n$
is infinite. Suppose the entanglement in the right $m$ pairs is $mE,$ where $%
E$ is average entanglement of right $m$ pairs unmeasured particles, after
the distinguishing measurement one get entanglement $mE$ from $n$ pairs. So
according to the definition of distillable entanglement\cite{6,12}

\begin{equation}
E_D=\frac{mE}n
\end{equation}
and E.q (\ref{9}) one can get the distillable entanglement of $\sigma $

\begin{equation}
E_D(\sigma )=(1-\frac{S(\sigma )}{I_{d\max }(\sigma )})E  \label{10}
\end{equation}
From E.q(\ref{10}) one can know that distillable entanglement is dependent
on $S(\sigma ),I_{d\max }(\sigma )$ and $E.$ E.q (\ref{10}) show the
relation of distinguishing information and distillable entanglement, and is
fit to any mixed state including multipartite mixed states. For general
mixed state it may be difficult to calculate $I_{d\max }(\sigma )$ and $E,$
in the following we discuss the case of Bell-diagonal states.

A Bell-diagonal state has following features:

1. Each one of the $2^{nS(\sigma )}$ ''likely'' strings in $\sigma ^{\otimes
n}$ is a maximally entangled state in $2^n\otimes 2^{n\text{ }}$ dimension
Hilbert space.

2. Any local unitary transformation on $\sigma ^{\otimes n}$ does not change
the entanglement of states, so after local unitary transformation each pair
particle in each string is still a Bell state.

3. For n copies of a Bell-diagonal state, $\rho ^{\otimes n},$ the
distinguishing measurements on each pair particle can only distinguish $%
\left| \Phi ^{\pm }\right\rangle $ from $\left| \Psi ^{\pm }\right\rangle ,$
and get the distinguishing information less or equal to 1 bit. This is
because for a Bell-diagonal state $\rho $, one at most can distinguish
locally two states of four Bell states under any LOCC operation\cite{2}. It
is to say that after local distinguishing one can distinguish 4 Bell states
into two group states. So the MDI is 1 bit, i.e.,

\begin{equation}
I_d(\rho )\leqslant 1\text{ bit}  \label{6}
\end{equation}
The equality holds if and only if two group states have same probability $%
\frac 12$. So the MDI of Bell-diagonal states is 1 bit.

Now we will show that there exist a set of local operation so that one can
get 1 bit MDI by the distinguishing measurements on an pair particle of the $%
2^{nS(\rho )}$ ''likely'' strings for n-copies of Bell-diagonal states $\rho
^{\otimes n}$. First we apply $(n-1)$ times bilateral Controlled-Not
(BCNOT)\ operation by letting the left first pair particle as the ''target''
and the second pair, third pair, ..., n'st pair as a ''source'' in turn. The
result of applying a BCNOT is\cite{9}: if the ''source'' is the state $%
\left| \Phi \right\rangle $, the ''target'' will unchange $\left| \Phi
\right\rangle $ or $\left| \Psi \right\rangle $; if the ''resource'' is the
state $\left| \Psi \right\rangle $, the ''target'' will exchange $\left|
\Phi \right\rangle \leftrightarrow \left| \Psi \right\rangle .$ The
''source'' pair will unchange $\left| \Phi \right\rangle $ or $\left| \Psi
\right\rangle $ in both cases. If the $n-1$ ''source'' pairs have odd (even) 
$\left| \Psi \right\rangle ,$ the ''target'' will exchange (unchange) $%
\left| \Phi \right\rangle \leftrightarrow \left| \Psi \right\rangle .$
Obviously when n is infinite the probability that the ''source'' has odd or
even $\left| \Psi \right\rangle $ is $\frac 12$ . So after $(n-1)$ times
BCNOT the probability that the ''target'' is $\left| \Phi \right\rangle $ or 
$\left| \Psi \right\rangle $ is exactly $\frac 12.$ Then one use projecting
operations, $\left| 00\right\rangle \left\langle 00\right| ,\left|
01\right\rangle \left\langle 01\right| ,\left| 10\right\rangle \left\langle
10\right| ,\left| 11\right\rangle \left\langle 11\right| ,$ on the first
pair and get 1 bit distinguishing information. Similarly we operate a set of
BCNOT by letting the $i$'st pair as ''target'' and j'st pair as ''source'', $%
i=2,...,(n-m),j=(i+1),(i+2),...,n.$ After we finish these operation each one
of the left $(n-m)$ pair particles has $\left| \Phi \right\rangle $ or $%
\left| \Psi \right\rangle $ with same probability $\frac 12.$ Thus we can
get $(n-m)$ bits information by measurement left $(n-m)$ pairs particles. So
for Bell-diagonal states the $I_{d\max }$ of each pair particle in the $%
2^{nS(\rho )}$ ''likely'' strings of $\rho ^{\otimes n}$ is 1 bit, and the $E
$ in E.q(\ref{10}) equal to 1. Thus we get the distillable entanglement of
Bell-diagonal state $\rho $ 

\begin{equation}
E_D(\rho )=1-S(\rho )  \label{11}
\end{equation}

It is well known that \cite{9} $1-S(\rho )$ is a lower bound of distillable
entanglement $E_D(\rho )$ of Bell-diagonal states. After consider the limit
of distinguishing locally and the relation between distillation and
distinguishing, we prove that $1-S(\rho )$ is also a upper bound of
distillable entanglement $E_D(\rho ).$ E.q(\ref{11}) imply that there are
some Bell-diagonal states, the information entropy of which $S(\rho )$ are
greater than unity, are entangled but have zero $E_D(\rho ).$ This is not
contradict with the claim \cite{13} that any entangled state in two-qubit
system is distillable. Because a state W is distillable imply that one can
get at least a maximally entangled state from infinite copies of the state
W, otherwise, W is a ''bound'' entangled state \cite{14}. However, only when
one get the same order infinite number of maximally entangled states from
infinite copies of the state W the $E_D(W)$ is not zero.

In summary, the transformation of information in distillation of
entanglement and distinguishability locally of quantum states act as a
important role. The relation of maximal distinguishing information and
distillable entanglement in E.q (\ref{10}) is fit to any mixed state. This
relation may be useful to calculate distillable entanglement and understand
the essence of entanglement. Although we only get the distillable
entanglement of a class of states here, we believer that this relation is
powerful in some others, especially in multi-partite and higher dimension of
bipartite system. A open question is how to calculate the maximal
distinguishing information of any mixed state.


\begin{references}
\bibitem{1}  C.H. Bennett, D.P. DiVincenzo, C.A. Fuchs, T.Mor, E.Rains, P.W.
Shor, J.A. Smolin, and W.K. Wootters, Phys. Rev. A 59,1070(1999) or
quant-ph/9804053.

\bibitem{2}  S.Ghosh, G.Kar, A.Roy, A.Sen and U.Sen, Phys.Rev.Lett.87,
277902 (2001)

\bibitem{3}  J.Walgate, A.J.Short, L.Hardy and V.Vedral,
Phys.Rev.Lett.85,4972(2000)

\bibitem{4}  Y.-X.Chen and D.Yang, Phys.Rev.A 64, 064303(2001)

\bibitem{5}  M. Horodecki, P. Horodecki, and R. Horodecki, Acta Physica
Slovaca, 48, (1998) 141, or quant-ph/9805072

\bibitem{6}  C. H. Bennett, D. P. Divincenzo, J. A.Smolin, and W.
K.Wootters, Phys. Rev. A{\bf \ 54}, 3824(1996).

\bibitem{7}  C.H. Bennett and S.J. Wiesner, Phys.Rev.Lett.69,2881(1992).

\bibitem{8}  C.H. Bennett, G.Brassard, C.Crepeau, R.Jozsa, A.Peres and
W.K.Wootters, Phys.Rev.Lett.70,1895(1993)

\bibitem{9}  C. H. Bennett, G. Brassard, S. Popescu, B. Schumacher, J. A.
Smolin, and W. K. Wootters, Phys. Rev. Lett{\bf \ 76} 722(1996).

\bibitem{10}  L.Henderson and V.Vedral, Phys.Rev.Lett84, 2263(2000).

\bibitem{11}  C.Brukner, M.Zukowski and A.Zeilinger, quant-ph/0106119

\bibitem{12}  E.M.Rains, Phys.Rew.A 60,173 (1999)

\bibitem{13}  M. Horodecki, P. Horodecki, and R. Horodecki, Phys.Rev.
Lett.78, 574(1997)

\bibitem{14}  M. Horodecki, P. Horodecki, and R. Horodecki, Phys.Rev.
Lett.80, 5239(1998)

\newpage
\end{references}
\end{document}